\documentclass[%
 amsmath,amssymb,
 aps,
prd,showpacs,onecolumn,superscriptaddress
]{revtex4-2}

\usepackage{graphicx}
\usepackage{subfigure}
\usepackage{dcolumn}

\usepackage{amsmath}
\usepackage{amssymb}
\usepackage{bm}
\usepackage{color}

\usepackage[normalem]{ulem}
\usepackage[dvipsnames]{xcolor}
\usepackage{hyperref}
\usepackage{bm}

\begin{document}

\preprint{APS/123-QED}

\title{Resolving ``dirty" effects around black holes by decoupling the Teukolsky equation}

\author{Ye Jiang}
 \affiliation{Shanghai Astronomical Observatory, Shanghai, 200030, China}
\affiliation{School of Astronomy and Space Science, University of Chinese Academy of Sciences,
Beijing, 100049, China}
\author{Wen-Biao Han}%
 \email{wbhan@shao.ac.cn}
\affiliation{School of Fundamental Physics and Mathematical Sciences, Hangzhou Institute for Advanced Study, UCAS, Hangzhou 310024, China}
\affiliation{Shanghai Astronomical Observatory, Shanghai, 200030, China}
\affiliation{School of Astronomy and Space Science, University of Chinese Academy of Sciences,
Beijing, 100049, China}
\affiliation{Shanghai Frontiers Science Center for  Gravitational Wave Detection, 800 Dongchuan Road, Shanghai 200240, China}
\date{\today}

\begin{abstract}
Detecting the environment around the supermassive black holes and tests of general relativity are important applications of extreme-mass-ratio inspirals (EMRIs). There is still a challenge to efficiently describe various ``dirty" impacts on the inspirals like dark matter, gas, dipole radiation, electromagnetic interaction, and so on. In this Letter, we find the inherent linearity of the asymptotic solution of the inhomogeneous Teukolsky equation. Based on this property, we completely decouple the factors of the perturber and the background spacetime in the energy fluxes and waveforms. With the new decoupling form, the waveforms of EMRIs with non-geodesic motion in Kerr spacetime can be calculated conveniently. This will help to resolve the environment (including gas, field, dark matter, electromagnetic interaction, etc.) around the supermassive black holes and test general relativity.
\end{abstract}

\maketitle


\noindent{{\bf{\em Introduction.}}} 
The discovery of gravitational waves (GWs) provides a new means of astronomical observation. LIGO, VIRGO, and KAGRA Collaborations (LVK) have observed more than one hundred GW events from comparable mass-ratio compact binaries in their sensitive frequency range ($10\;\rm{Hz}-10^3\;\rm{Hz}$) \citep{PhysRevLett.116.061102, theligoscientificcollaboration2021gwtc3}. Space-borne gravitational wave detectors like Laser interferometer space antenna (LISA) and Taiji are expected to detect the gravitational wave lying in the milli-Hertz band \citep{amaroseoane2017laser, 10.1093/nsr/nwx116}. The potential of these space-borne detectors extends to capturing GWs emanating from a diverse array of astrophysical and cosmological sources \cite{Amaro-Seoane2023}.

Extreme mass-ratio inspirals (EMRI) are among the most interesting GW sources for upcoming space-based GW detectors. In an EMRI, a stellar secondary of mass $\mu$ evolves around a supermassive black hole (SMBH) of mass $M$ with mass ratio $q=\mu/M\sim 10^{-4}-10^{-7}$, and the secondary object completes approximately $\mathcal{O}(1/q)$ orbits around the primary before its final plunge \citep{PhysRevD.95.103012}. In particular, the EMRIs are expected to offer unprecedented precision in parameter estimation for astrophysics \citep{AmaroSeoane2021, PhysRevLett.129.241103, PhysRevD.107.024035} and fundamental physics \citep{Barack_2019, Barausse2020, PhysRevD.107.024021}.

Detecting or constraining the parameters of the EMRIs requires accurate waveform templates. The primary method of modeling EMRIs is the black hole perturbation theory, which solves the Einstein field equations using an expansion in powers of the mass ratio $q$ \citep{PhysRevD.78.064028, PhysRevD.106.124040}. Early work in perturbation theory considers the secondary as a test particle in vacuum environments \cite{10.1143/PTPS.128.1}. In actuality, EMRI events can form in a variety of interesting astrophysical environments and interact with materials in the environment, e.g., gas \citep{10.1093/mnras/stad3132, 2023A&A...675A.100F}, dark matter \cite{Cole2023}, scale field \citep{PhysRevLett.125.141101, PhysRevLett.131.051401}, vector field \citep{Zhang_2023, PhysRevD.107.044053}, and etc. To conceptually distinguish with the usual EMRIs, in this Letter we call this kind of systems as ``dirty" EMRIs due to the complexities. Recent researches also took the inner structure of the secondary into account \citep{PhysRevD.107.024006, PhysRevD.102.024041}.

These environmental or non-gravitational influences will induce various effects on the GW signals. Dynamic friction can be caused by gas or dark matter \citep{PhysRevLett.110.221101, Li2022}. If the charged stellar mass secondary moves in a field, that will introduce an extra dissipation beside the GW radiation \citep{PhysRevLett.125.141101, PhysRevD.108.084019}. The inner structure of the secondary affects its dynamical features which are described by the Mathisson-Papapetrou-Dixon (MPD) equation \cite{PhysRevD.81.044019}. Even the mass of the secondary can also cause a non-negligible effect for testing general relativity \cite{PhysRevD.108.064015}. Detecting those effects from GWs will provide unique measurements of environments around MBHs through GW observations and impose constraints on gas densities, dark matter profiles, or the presence of external perturbations.

However, the diversity and complexity of the environment presents computational difficulties of the waveform, there is no general waveform model to describe those varied situations. In this Letter, to solve this problem, we present a new method that separates the secondary's extrinsic influences and intrinsic structure from the GW energy flux in the circular orbit and the waveform for the quasicircular inspiral. Our method greatly simplifies the waveform calculation and brings various effects into the same framework.

\noindent{{\bf{\em Decoupling form and analysis.}}}
Currently, the accurate calculation of EMRI waveforms is based on the Teukolsky formalism \citep{1973ApJ...185..635T, 1973ApJ...185..649P, 1974ApJ...193..443T} in which the perturbation of the Kerr background is decomposed by the Newman-Penrose tetrad basis. 
At infinity, the two GW polarizations are both encoded in the $\Psi_4$ Weyl scalar:
\begin{equation}
    \Psi_4(r\rightarrow\infty)=\frac{1}{2}\frac{\partial^2}{\partial t^2}(h_+-ih_\times).
\end{equation}

Teukolsky showed that the $\Psi_4$ has the following multipolar decomposition \cite{1973ApJ...185..635T}
\begin{equation}
    \Psi_4=\rho^4\sum_{l=2}^{\infty}\sum_{m=-l}^l\int_{-\infty}^{\infty}d\omega R_{lm\omega}(r)S_{lm}^{a\omega}(\theta)e^{i(m\phi-\omega t)},
\end{equation}
where $\rho=(r-ia\cos{\theta})^{-1}$ and $S_{lm}^{a\omega}(\theta)$ is the spin weighted spheroidal harmonics with weight $-2$. 

The radial function $R_{lm\omega}$ obeys the Teukolsky equation (master equation) 
\begin{equation}
    \Delta^2\frac{d}{dr}\left(\frac{1}{\Delta}\frac{dR_{lm\omega}}{dr}\right)-V(r)R_{lm\omega}(r)=\mathcal{T}_{lm\omega},
\end{equation}
where the source term $\mathcal{T}$ is decided by the stress-energy tensor of the perturber (the secondary compact object). $R_{lm\omega}$ has the asymptotic expression at horizon and infinity
\begin{align}
    R_{lm\omega}(r\rightarrow r_+)=Z^{\infty}_{lm\omega}\Delta^2e^{-i\kappa r^*},\\
    R_{lm\omega}(r\rightarrow \infty)=Z^{H}_{lm\omega}r^3e^{-i\omega r^*},
\end{align}
where $\kappa=\omega-ma/2Mr_+$, $r_+=M+\sqrt{M^2-a^2}$.

There are two issues for calculating the waveforms of ``dirty" EMRIs. One is due to the dynamical effects, the orbits of the secondary objects will be deviated from the shrinking geodesic and then induce changes of the frequency and evolution of gravitational waves. This may be calculated easily. Secondly, the deviation of the motion will also change the perturbation itself on the background of the supermassive black holes. Due to the ``dirty" effects, the source term become much complicated than the test particle approximation. As a result, the properties of the perturber couple with the background spacetime and make solving the Teukolsky radial equation to be very difficult even impossible.  

The asymptotic amplitude $Z^{H,\infty}_{lm\omega}$ is related to the stress-energy tensor $T^{\alpha\beta}$ of the perturbation source. Usually, in the calculation of EMRI waveforms, $T^{\alpha\beta}$ is obtained by the test particle approximation which omits the intrinsic structure and in a ``pure'' Kerr spacetime. However, firstly the secondary compact object should be an extended body; Secondly, in the vicinity of SMBH there may be dark matter halos, accretion gas, and so on; Finally, scalar field, dipole radiation, and electromagnetic interaction may exist. All these effects could induce a deviation of the motion of the secondary body from the adiabatic orbital evolution of the test particle. This is why we call the EMRIs in some of these situations as ``dirty" EMRIs. To address this challenge and establish a general template for various dirty EMRIs, we meticulously analyze the structure of $Z^{H,\infty}_{lm\omega}$ and find out its inherent linearity. 

This finding can be represented as follows. Based on ~\cite{PhysRevD.81.044019}, the stress-energy tensor of the compact object has the general form,
\begin{equation}\label{eq: stress-energy tensor}
    T^{\alpha\beta}=\int \left\{t^{\alpha\beta}\delta_{(4)}+\nabla_\gamma[t^{\gamma\alpha\beta}\delta_{(4)}]+\nabla_\delta\nabla_\gamma[t^{\delta\gamma\alpha\beta}\delta_{(4)}]+ ... \right\}d\tau,
\end{equation}
where $\delta_{(4)}=\delta(x-z(\tau))/\sqrt{-g}$, $z(\tau)$ is the worldline of the particle. ($t^{\alpha\beta}$, $t^{\alpha\beta\gamma}$, ...) were called as general multipole moments of the perturber \cite{tulczyjew1959motion}, now we also include the no-vacuum or no-gravitational effects like as gas, electromagnetic interaction, scalar field, dipole radiation and so on. This arises from the fact that, for arbitrary tensor singular along the worldline, this general form always holds \cite{PhysRevD.81.044019}. And if such a compact object moves in an equatorial circular orbit around the SMBH, then the asymptotic amplitude $Z^{H,\infty}_{lm\omega}=2\pi \mathcal{A}^{H,\infty}_{lm}\delta(\omega-m\Omega)$, where $\Omega$ is the orbit frequency. In the previous literature, $\mathcal{A}^{H,\infty}_{lm}$ mix the information of secondary object and the background gravitational field, and will be too complicated to solve the master equation for dirty EMRIs. Now, for the first time, we find that it can be written in the form of 
\begin{equation}
    \mathcal{A}^{H,\infty}_{lm}=\sum_ix_iC_{ilm}^{in,up}(a,\omega,r),\label{eq: decoupling amplitude}
\end{equation}
where the coefficients $x_i$ describe properties of the perturbation source. $x_i$ are given from components of the general multipole moments divided by $v^t$, and in principle are required independent of $t$. If the $x_i$ slowly changes with time, then the above equation also holds with adiabatic approximation. $C_{ilm}^{in, up}$ are linear combination coefficients only decided by the Kerr spacetime, and $a$ is the spin of the SMBH. The proof of Eq.~(\ref{eq: decoupling amplitude}) can be found in supplementary materials.

For a test particle along the geodesic trajectory, things are quite simple, particle's stress-energy tensor is $T^{\alpha\beta}=\int \mu v^{\alpha}v^{\beta}\delta_{(4)}d\tau$, then $x_i=\{\mu v^\phi \Omega, \mu v^t, \mu v^\phi\}$. In previous studies, the common coefficient $\mu$ is extracted from the asymptotic amplitude, avoiding the need to repetitively calculate $\mathcal{A}^{H,\infty}_{lm}$ for different $\mu$ values. However, for a compact object deviating from the geodesic trajectory (possibly caused by gas, field, dark matter, dipole radiation, etc.), the asymptotic amplitude given in the traditional method \cite{10.1143/PTPS.128.1} intertwines with parameters of secondary, which makes it troublesome to reevaluate $\mathcal{A}^{H,\infty}_{lm}$ for each particular perturbation source. Moreover, for some more complicated perturbation sources, an explicit expression for the asymptotic amplitude may be not available from the traditional method. For solving this problem, our well-structured formula Eq.~(\ref{eq: decoupling amplitude}) of $\mathcal{A}^{H,\infty}_{lm}$ allows us to completely separate parameters of the perturbation source from the background spacetime.

The total GWs' energy flux can be written as the sum of the flux at infinity and horizon:
\begin{equation}\label{eq: flux}
    \mathcal{F}_{\text{GW}}=\mathcal{F}^\infty+\mathcal{F}^{H}=\sum_{lm}\left.\frac{\left|\mathcal{A}^H_{lm}\right|^2}{2\pi\omega^2}\right|_{\omega=m\Omega}+\sum_{lm}\left.\frac{\alpha_{lm\omega}\left|\mathcal{A}^\infty_{lm}\right|^2}{2\pi\omega^2}\right|_{\omega=m\Omega},
\end{equation}
where $\mathcal{F}^\infty$ is calculated from the Isaacson stress-energy tensor \cite{PhysRev.166.1272}, $\mathcal{F}^H$ is obtained by measuring the rate at which the event horizon’s area increases as radiation falls into it, and the coefficient $\alpha_{lm\omega}$ is determined through a transformation from Kinnerley’s null tetrad (used to construct $\psi_4$) to the Hawking-Hartle null tetrad \citep{Hawking1972, 1974ApJ...193..443T}.

Here, to remove the square of the norm $\left|\cdot\right|^2$ in Eq.~(\ref{eq: flux}), we introduce a simple lemma for the following simplification. If $a_i$ are complex numbers, $x_i$ are real numbers and $i\in I$ a countable index set, then we have
\begin{equation}
    \left|\sum_{i\in I}a_ix_i\right|^2=\sum_{i\in I}\sum_{j\in I}x_ix_j(a_i\diamond a_j),
\end{equation}
where the operator $\diamond$ is defined by
\begin{equation}
    a\diamond b=\rm{Re}(a)\rm{Re}(b)+\rm{Im}(a)\rm{Im}(b).
\end{equation}

We then convert the square of the norm in Eq.~(\ref{eq: flux}) to a sum, and switch the order of the sum to get the following new form of GW energy flux:
\begin{equation}\label{eq: decoupling flux}
    \mathcal{F}_{\text{GW}}=\sum_{ij}x_ix_jf_{ij}(a,\Omega,r)
\end{equation}
where
\begin{equation}
    f_{ij}=2\pi\sum_{l=2}^{\infty}\sum_{m=1}^{l}\left.((C^{in}_{ilm}\diamond C^{in}_{jlm})+\alpha_{lm}(C^{up}_{ilm}\diamond C^{up}_{jlm}))/\omega^2\right|_{\omega=m\Omega}
\end{equation}
are functions only decided by the Kerr spacetime. Our well-structured formula Eq.~(\ref{eq: decoupling flux}) of $\mathcal{F}_{\text{GW}}$ allows us to completely separate the perturber's properties from the background. Consequently, when computing the GW energy flux for the ``dirty" EMRIs, $f_{ij}$ does not relate with the complicated situation, is just obtained from the Kerr spacetime properties, and in principle can be fitted by a three-dimensional ($a,~ r, ~\omega$) interpolation. Therefore the energy fluxes for ``dirty" EMRIs with varied scenarios can be simply calculated by multiply the source's coefficients $x_i$. Additionally, the angular momentum flux at infinity or horizon also has a similar form with Eq.~(\ref{eq: decoupling flux}).

Based on the adiabatic approximation, we could determine an equation for $dr/dt$ by employing an energy balance law, $dr/dt=-(\mathcal{F}_{\rm{GW}}+\mathcal{F}_{\rm{D}})(dE/dr)^{-1}$, where $E$ is the orbit energy and $\mathcal{F}_{\rm{D}}$ is dissipation caused by the dirty effects. Thus, we derive a new form of the quasicircular inspiral waveform by applying the Eq.~(\ref{eq: decoupling amplitude}),
\begin{equation}\label{eq: decoupling waveform}
    h_+-ih_\times=-\frac{2}{D}\sum_ix_i[r(t)]\sum_{l=2}^{\infty}\sum_{m=-l}^{l}\left\{C^{\text{in}}_{ilm}[a,m\Omega(t),r(t)]S_{lm}(\Theta)e^{im[\Phi-\phi(t)]}\right\},
\end{equation}
where $D$ is luminosity distance, $\Theta$ is the angle between an observer's line of sight and the primary's spin axis, and $\Phi$ is the azimuthal angle. Similarly, our new formula Eq.~(\ref{eq: decoupling waveform}) of the waveform has same advantage with the fluxes. Our decoupling formula can be used to develop a fast model-independent waveform template for dirty EMRIs.


To show the potential of analyzing dirty EMRIs with our method, we then consider a system composed of a $10^6M_\odot$ SMBH and a $10M_\odot$ secondary object in some certain astrophysical environments, $M_\odot$ is the solar mass. To simplify the analysis but without loss of generality, we assume the stress-energy tensor of the secondary takes the following form
\begin{equation}
     T^{\alpha\beta}=\int p^{\alpha}v^{\beta}\delta_{(4)}d\tau,
\end{equation}
where $p^{\alpha}=\mu v^{\alpha}$, $v^{\beta}=v^{\beta}_\text{D}+v^{\beta}_\text{V}$, $v^{\beta}_\text{V}$ is the 4-velocity of the secondary with test particle approximation in the vacuum and $v^{\beta}_\text{D}$ is an arbitrary correction due to the ``dirty" effect (possibly caused by gas, field, dark matter, dipole radiation, etc.), other quantities have the same definition with Eq.~(\ref{eq: stress-energy tensor}). We then calculate the one-year duration EMRI waveforms stop at the innermost stable circular orbit (ISCO). 
\begin{figure}[htbp!]
    \centering
    
    \begin{minipage}{0.49\linewidth}
        \centering
        \includegraphics[width=\linewidth]{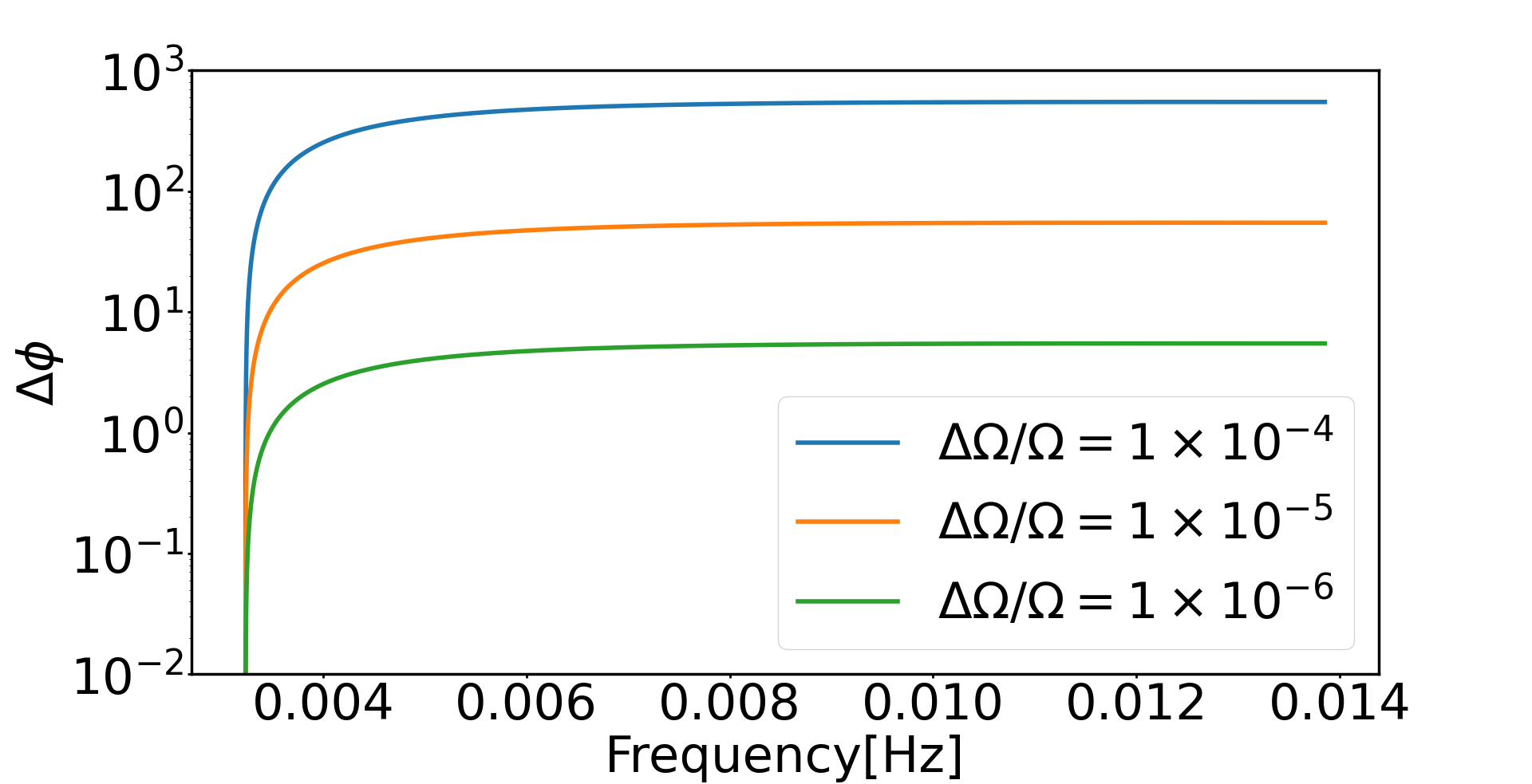}
    \end{minipage}
    \begin{minipage}{0.49\linewidth}
        \centering
        \includegraphics[width=\linewidth]{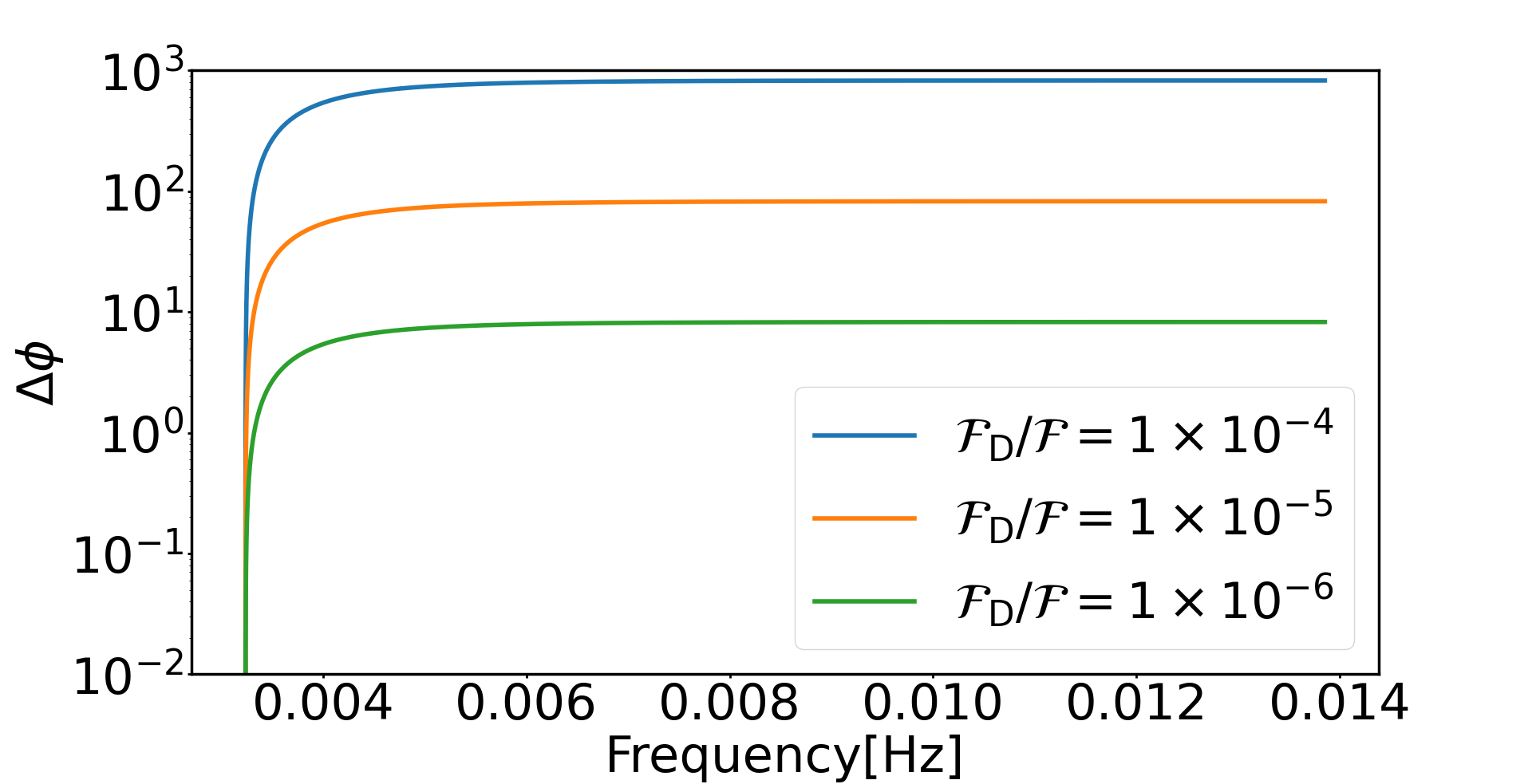}
    \end{minipage}

    \caption{The dephase $\Delta\phi$ of the ``dirty"  EMRI waveforms as a function of frequency for quasicircular, prograde orbits, $\hat{a}=0.9$. Here, $\hat{a}=a/M$ is the dimensionless spin of SMBH, $\Delta\Omega=v^{\phi}_\text{D}/v^{t}$ is the orbit frequency deviation caused by ``dirty" effect.} 
    \label{fig: phase frequency}
\end{figure}

We first consider the phase deviation caused by the ``dirty"  effect in the frequency domain. Two waveforms would be indistinguishable through the measurement of the detector if the phase deviation $\Delta\phi>\sqrt{D-1}/\text{SNR}$, where $D$ is the number of intrinsic source parameters in the waveform, and SNR is the measured signal-to-noise ratio. We typically choose $\text{SNR}=20$, following Ref.~\cite{PhysRevLett.123.101103}. Thus, one can roughly estimate that the phase resolution is $\Delta\phi\sim 0.1$. Fig.~\ref{fig: phase frequency} shows the dephasing of waveforms due to the $\Delta\Omega$ and $\mathcal{F}_\text{D}$ which are both caused by the ``dirty" effects. We can see that future space-borne detectors can resolve the ``dirty" effect with high accuracy, measure the relative deviation from orbital frequency and flux at a level $\sim 10^{-6}$. 

To further quantify the detection potential of ``dirty" EMRI, we employ the Fisher matrix to evaluate the parameter estimate accuracy. For a large SNR, the posterior distribution of $\theta$ inferred by an EMRI detection can be approximated by a Gaussian centered around the true values $\theta_t$, with covariance $\mathbf{\Sigma}=\mathbf{\Gamma}^{-1}$, where $\mathbf{\Gamma}$ is the Fisher information matrix. The square of the root of diagonal elements $\Sigma_{ii}^{1/2}$ corresponds to the statistical error $\Delta \theta_i$ of the $i$th parameter \cite{PhysRevD.77.042001}.

\begin{figure}[htbp!]
    \centering
    
    \begin{minipage}{0.45\linewidth}
        \centering
        \includegraphics[width=\linewidth]{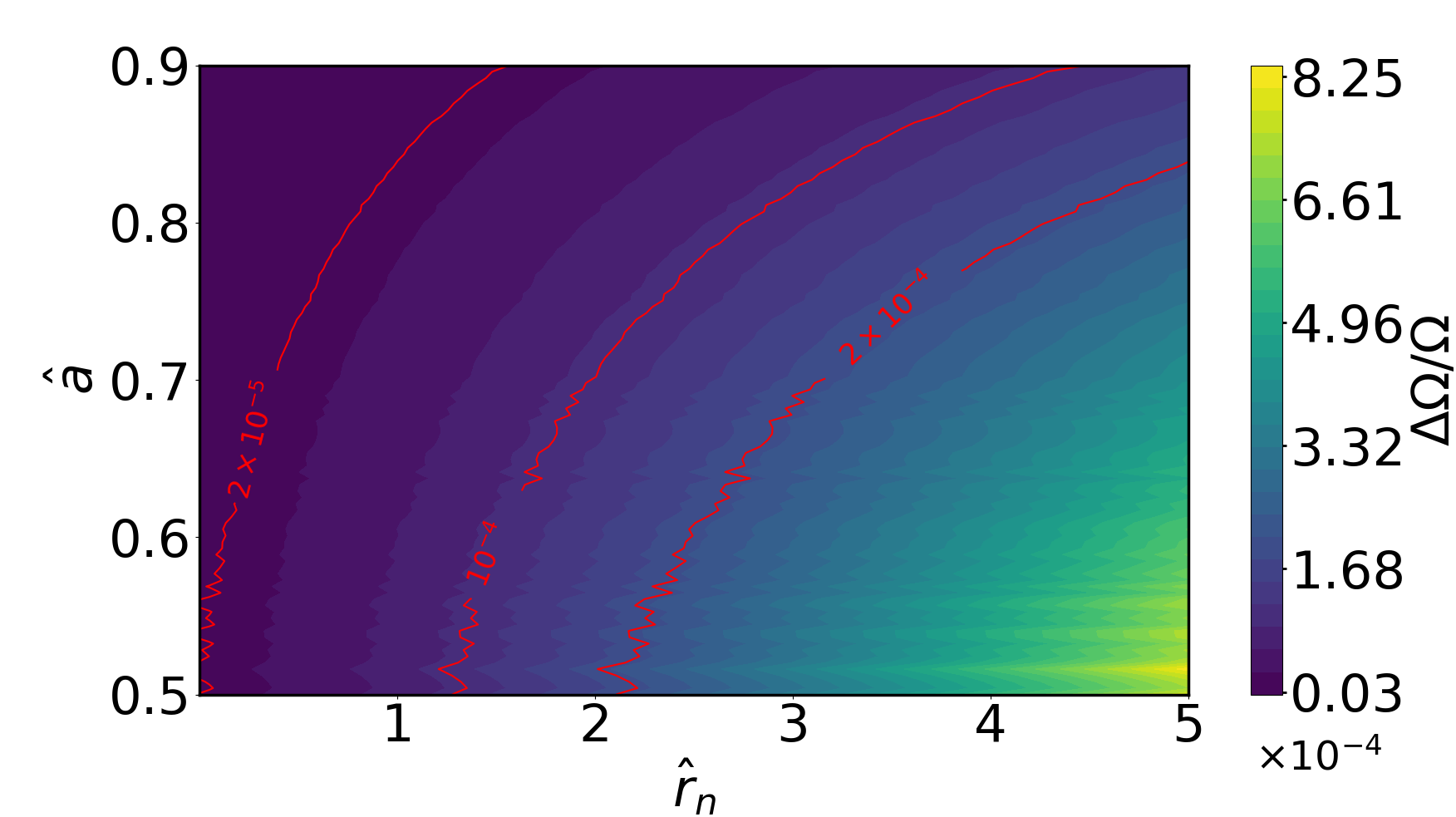}
    \end{minipage}
    \begin{minipage}{0.45\linewidth}
        \centering
        \includegraphics[width=\linewidth]{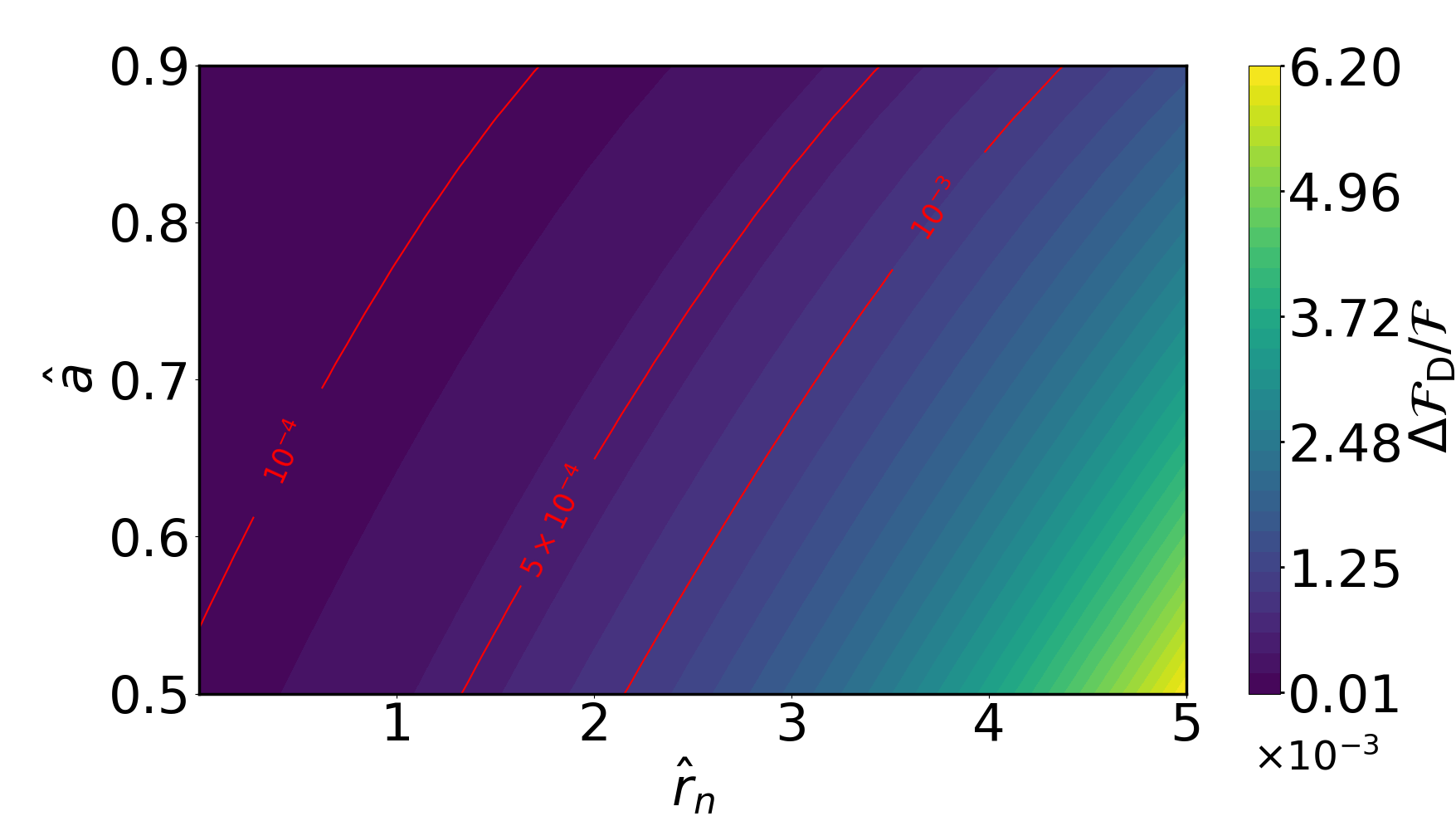}
    \end{minipage}

    \caption{The relative error of $\Omega$ (left panel)and $\mathcal{F}_\text{D}$ (right panel) for the distribution of $\hat{a}$ and $\hat{r}_n$, where $\mathcal{F}=\mathcal{F}_\text{D}+\mathcal{F}_\text{V}$ is the total energy flux, $\hat{r}_n=(r-r_{\text{ISCO}})/M$ is the dimensionless distance from $r_\text{ISCO}$.}
    \label{fig: countor_fisher}
\end{figure}

The results of the parameter estimate accuracy are presented in Fig.~\ref{fig: countor_fisher}. The $\Delta\Omega$ in the left panel of Fig.~\ref{fig: countor_fisher} are calculated by the probability density distribution (PDF) of $(v^{t}_\text{D},v^{\phi}_\text{D})$ which is a norm distribution with the covariance obtained from the Fisher matrix. $\Delta\mathcal{F}_{\text{D}}$ in the left panel of Fig.~\ref{fig: countor_fisher} is obtained directly through the corresponding diagonal element in the $\Sigma$.

From Fig.~\ref{fig: countor_fisher}, the deviation of the orbit frequency and energy flux caused by the ``dirty" effect can be detected with high accuracy, the relative error of $\Omega$ and $\mathcal{F}$ are as small as $\sim 10^{-6}$ and $\sim 10^{-5}$, respectively. Such results demonstrate the extraordinary potential for detecting ``dirty" effects by future space-borne GW detectors. 

\noindent{{\bf{\em Conclusion.}}}
Resolving the non-vacuum environments or non-GR effects around SMBHs are key science targets of EMRIs. However, there is no general waveform model to describe the varied situations. In this Letter, we decompose the asymptotic amplitude of the Teukolsky formalism and can accurately calculate the waveforms for arbitrary effects that induce derivations from the vacuum and GR situation, e.g., dynamic friction from the environment (dark matter or gas), dipole radiation, and electromagnetic interaction. A model-independent waveform template could be developed based on our new method.

Eq.~(\ref{eq: decoupling amplitude}) indicate the inherent linearity of the asymptotic amplitude which revealed the structure of perturbations of black holes. Base on it, our new forms, given by Eq.~(\ref{eq: decoupling flux}) and Eq.~(\ref{eq: decoupling waveform}) describe the GW energy flux and the waveform with separated secondary’s factors. Similar results could be derived for different $s$ in the future, where $s$ represents the spin weight of the field, $s=0$ for the scalar field, $s=\pm1$ for the vector field, $s=\pm2$ for a gravitational perturbation. For another possible extension of this work, generalizing the result to the generic orbits is attractive, but one may encounter some technical challenges. 


\begin{thebibliography}{38}%
\makeatletter
\providecommand \@ifxundefined [1]{%
 \@ifx{#1\undefined}
}%
\providecommand \@ifnum [1]{%
 \ifnum #1\expandafter \@firstoftwo
 \else \expandafter \@secondoftwo
 \fi
}%
\providecommand \@ifx [1]{%
 \ifx #1\expandafter \@firstoftwo
 \else \expandafter \@secondoftwo
 \fi
}%
\providecommand \natexlab [1]{#1}%
\providecommand \enquote  [1]{``#1''}%
\providecommand \bibnamefont  [1]{#1}%
\providecommand \bibfnamefont [1]{#1}%
\providecommand \citenamefont [1]{#1}%
\providecommand \href@noop [0]{\@secondoftwo}%
\providecommand \href [0]{\begingroup \@sanitize@url \@href}%
\providecommand \@href[1]{\@@startlink{#1}\@@href}%
\providecommand \@@href[1]{\endgroup#1\@@endlink}%
\providecommand \@sanitize@url [0]{\catcode `\\12\catcode `\$12\catcode
  `\&12\catcode `\#12\catcode `\^12\catcode `\_12\catcode `\%12\relax}%
\providecommand \@@startlink[1]{}%
\providecommand \@@endlink[0]{}%
\providecommand \url  [0]{\begingroup\@sanitize@url \@url }%
\providecommand \@url [1]{\endgroup\@href {#1}{\urlprefix }}%
\providecommand \urlprefix  [0]{URL }%
\providecommand \Eprint [0]{\href }%
\providecommand \doibase [0]{https://doi.org/}%
\providecommand \selectlanguage [0]{\@gobble}%
\providecommand \bibinfo  [0]{\@secondoftwo}%
\providecommand \bibfield  [0]{\@secondoftwo}%
\providecommand \translation [1]{[#1]}%
\providecommand \BibitemOpen [0]{}%
\providecommand \bibitemStop [0]{}%
\providecommand \bibitemNoStop [0]{.\EOS\space}%
\providecommand \EOS [0]{\spacefactor3000\relax}%
\providecommand \BibitemShut  [1]{\csname bibitem#1\endcsname}%
\let\auto@bib@innerbib\@empty
\bibitem [{\citenamefont {Abbott}\ \emph {et~al.}(2016)\citenamefont {Abbott},
  \citenamefont {Abbott} \emph {et~al.}}]{PhysRevLett.116.061102}%
  \BibitemOpen
  \bibfield  {author} {\bibinfo {author} {\bibfnamefont {B.~P.}\ \bibnamefont
  {Abbott}}, \bibinfo {author} {\bibfnamefont {R.}~\bibnamefont {Abbott}},
  \emph {et~al.} (\bibinfo {collaboration} {LIGO Scientific Collaboration and
  Virgo Collaboration}),\ }\bibfield  {title} {\bibinfo {title} {Observation of
  gravitational waves from a binary black hole merger},\ }\href
  {https://doi.org/10.1103/PhysRevLett.116.061102} {\bibfield  {journal}
  {\bibinfo  {journal} {Phys. Rev. Lett.}\ }\textbf {\bibinfo {volume} {116}},\
  \bibinfo {pages} {061102} (\bibinfo {year} {2016})}\BibitemShut {NoStop}%
\bibitem [{\citenamefont {Abbott}\ \emph {et~al.}(2021)\citenamefont {Abbott}
  \emph {et~al.}}]{theligoscientificcollaboration2021gwtc3}%
  \BibitemOpen
  \bibfield  {author} {\bibinfo {author} {\bibfnamefont {R.}~\bibnamefont
  {Abbott}} \emph {et~al.} (\bibinfo {collaboration} {LIGO Scientific, VIRGO,
  and KAGRA Collaborations}),\ }\href@noop {} {\bibinfo {title} {Gwtc-3:
  Compact binary coalescences observed by ligo and virgo during the second part
  of the third observing run}} (\bibinfo {year} {2021}),\ \Eprint
  {https://arxiv.org/abs/2111.03606} {arXiv:2111.03606 [gr-qc]} \BibitemShut
  {NoStop}%
\bibitem [{\citenamefont {Amaro-Seoane}\ \emph {et~al.}(2017)\citenamefont
  {Amaro-Seoane} \emph {et~al.}}]{amaroseoane2017laser}%
  \BibitemOpen
  \bibfield  {author} {\bibinfo {author} {\bibfnamefont {P.}~\bibnamefont
  {Amaro-Seoane}} \emph {et~al.} (\bibinfo {collaboration} {LISA
  Collaboration}),\ }\href@noop {} {\bibinfo {title} {Laser interferometer
  space antenna}} (\bibinfo {year} {2017}),\ \Eprint
  {https://arxiv.org/abs/1702.00786} {arXiv:1702.00786 [astro-ph.IM]}
  \BibitemShut {NoStop}%
\bibitem [{\citenamefont {Hu}\ and\ \citenamefont
  {Wu}(2017)}]{10.1093/nsr/nwx116}%
  \BibitemOpen
  \bibfield  {author} {\bibinfo {author} {\bibfnamefont {W.-R.}\ \bibnamefont
  {Hu}}\ and\ \bibinfo {author} {\bibfnamefont {Y.-L.}\ \bibnamefont {Wu}},\
  }\bibfield  {title} {\bibinfo {title} {{The Taiji Program in Space for
  gravitational wave physics and the nature of gravity}},\ }\href
  {https://doi.org/10.1093/nsr/nwx116} {\bibfield  {journal} {\bibinfo
  {journal} {National Science Review}\ }\textbf {\bibinfo {volume} {4}},\
  \bibinfo {pages} {685} (\bibinfo {year} {2017})},\ \Eprint
  {https://arxiv.org/abs/https://academic.oup.com/nsr/article-pdf/4/5/685/31566708/nwx116.pdf}
  {https://academic.oup.com/nsr/article-pdf/4/5/685/31566708/nwx116.pdf}
  \BibitemShut {NoStop}%
\bibitem [{\citenamefont {Amaro-Seoane}\ \emph {et~al.}(2023)\citenamefont
  {Amaro-Seoane} \emph {et~al.}}]{Amaro-Seoane2023}%
  \BibitemOpen
  \bibfield  {author} {\bibinfo {author} {\bibnamefont {Amaro-Seoane}} \emph
  {et~al.},\ }\bibfield  {title} {\bibinfo {title} {Astrophysics with the laser
  interferometer space antenna},\ }\href
  {https://doi.org/10.1007/s41114-022-00041-y} {\bibfield  {journal} {\bibinfo
  {journal} {Living Reviews in Relativity}\ }\textbf {\bibinfo {volume} {26}},\
  \bibinfo {pages} {2} (\bibinfo {year} {2023})}\BibitemShut {NoStop}%
\bibitem [{\citenamefont {Babak}\ \emph {et~al.}(2017)\citenamefont {Babak},
  \citenamefont {Gair}, \citenamefont {Sesana}, \citenamefont {Barausse},
  \citenamefont {Sopuerta}, \citenamefont {Berry}, \citenamefont {Berti},
  \citenamefont {Amaro-Seoane}, \citenamefont {Petiteau},\ and\ \citenamefont
  {Klein}}]{PhysRevD.95.103012}%
  \BibitemOpen
  \bibfield  {author} {\bibinfo {author} {\bibfnamefont {S.}~\bibnamefont
  {Babak}}, \bibinfo {author} {\bibfnamefont {J.}~\bibnamefont {Gair}},
  \bibinfo {author} {\bibfnamefont {A.}~\bibnamefont {Sesana}}, \bibinfo
  {author} {\bibfnamefont {E.}~\bibnamefont {Barausse}}, \bibinfo {author}
  {\bibfnamefont {C.~F.}\ \bibnamefont {Sopuerta}}, \bibinfo {author}
  {\bibfnamefont {C.~P.~L.}\ \bibnamefont {Berry}}, \bibinfo {author}
  {\bibfnamefont {E.}~\bibnamefont {Berti}}, \bibinfo {author} {\bibfnamefont
  {P.}~\bibnamefont {Amaro-Seoane}}, \bibinfo {author} {\bibfnamefont
  {A.}~\bibnamefont {Petiteau}},\ and\ \bibinfo {author} {\bibfnamefont
  {A.}~\bibnamefont {Klein}},\ }\bibfield  {title} {\bibinfo {title} {Science
  with the space-based interferometer lisa. v. extreme mass-ratio inspirals},\
  }\href {https://doi.org/10.1103/PhysRevD.95.103012} {\bibfield  {journal}
  {\bibinfo  {journal} {Phys. Rev. D}\ }\textbf {\bibinfo {volume} {95}},\
  \bibinfo {pages} {103012} (\bibinfo {year} {2017})}\BibitemShut {NoStop}%
\bibitem [{\citenamefont {Amaro~Seoane}\ \emph {et~al.}(2021)\citenamefont
  {Amaro~Seoane} \emph {et~al.}}]{AmaroSeoane2021}%
  \BibitemOpen
  \bibfield  {author} {\bibinfo {author} {\bibfnamefont {P.}~\bibnamefont
  {Amaro~Seoane}} \emph {et~al.},\ }\bibfield  {title} {\bibinfo {title} {The
  effect of mission duration on lisa science objectives},\ }\href
  {https://doi.org/10.1007/s10714-021-02889-x} {\bibfield  {journal} {\bibinfo
  {journal} {General Relativity and Gravitation}\ }\textbf {\bibinfo {volume}
  {54}},\ \bibinfo {pages} {3} (\bibinfo {year} {2021})}\BibitemShut {NoStop}%
\bibitem [{\citenamefont {Cardoso}\ \emph {et~al.}(2022)\citenamefont
  {Cardoso}, \citenamefont {Destounis}, \citenamefont {Duque}, \citenamefont
  {Macedo},\ and\ \citenamefont {Maselli}}]{PhysRevLett.129.241103}%
  \BibitemOpen
  \bibfield  {author} {\bibinfo {author} {\bibfnamefont {V.}~\bibnamefont
  {Cardoso}}, \bibinfo {author} {\bibfnamefont {K.}~\bibnamefont {Destounis}},
  \bibinfo {author} {\bibfnamefont {F.}~\bibnamefont {Duque}}, \bibinfo
  {author} {\bibfnamefont {R.~P.}\ \bibnamefont {Macedo}},\ and\ \bibinfo
  {author} {\bibfnamefont {A.}~\bibnamefont {Maselli}},\ }\bibfield  {title}
  {\bibinfo {title} {Gravitational waves from extreme-mass-ratio systems in
  astrophysical environments},\ }\href
  {https://doi.org/10.1103/PhysRevLett.129.241103} {\bibfield  {journal}
  {\bibinfo  {journal} {Phys. Rev. Lett.}\ }\textbf {\bibinfo {volume} {129}},\
  \bibinfo {pages} {241103} (\bibinfo {year} {2022})}\BibitemShut {NoStop}%
\bibitem [{\citenamefont {Bamber}\ \emph {et~al.}(2023)\citenamefont {Bamber},
  \citenamefont {Aurrekoetxea}, \citenamefont {Clough},\ and\ \citenamefont
  {Ferreira}}]{PhysRevD.107.024035}%
  \BibitemOpen
  \bibfield  {author} {\bibinfo {author} {\bibfnamefont {J.}~\bibnamefont
  {Bamber}}, \bibinfo {author} {\bibfnamefont {J.~C.}\ \bibnamefont
  {Aurrekoetxea}}, \bibinfo {author} {\bibfnamefont {K.}~\bibnamefont
  {Clough}},\ and\ \bibinfo {author} {\bibfnamefont {P.~G.}\ \bibnamefont
  {Ferreira}},\ }\bibfield  {title} {\bibinfo {title} {Black hole merger
  simulations in wave dark matter environments},\ }\href
  {https://doi.org/10.1103/PhysRevD.107.024035} {\bibfield  {journal} {\bibinfo
   {journal} {Phys. Rev. D}\ }\textbf {\bibinfo {volume} {107}},\ \bibinfo
  {pages} {024035} (\bibinfo {year} {2023})}\BibitemShut {NoStop}%
\bibitem [{\citenamefont {Askar}\ \emph {et~al.}(2019)\citenamefont {Askar},
  \citenamefont {Belczynski} \emph {et~al.}}]{Barack_2019}%
  \BibitemOpen
  \bibfield  {author} {\bibinfo {author} {\bibfnamefont {A.}~\bibnamefont
  {Askar}}, \bibinfo {author} {\bibfnamefont {C.}~\bibnamefont {Belczynski}},
  \emph {et~al.},\ }\bibfield  {title} {\bibinfo {title} {Black holes,
  gravitational waves and fundamental physics: a roadmap},\ }\href
  {https://doi.org/10.1088/1361-6382/ab0587} {\bibfield  {journal} {\bibinfo
  {journal} {Classical and Quantum Gravity}\ }\textbf {\bibinfo {volume}
  {36}},\ \bibinfo {pages} {143001} (\bibinfo {year} {2019})}\BibitemShut
  {NoStop}%
\bibitem [{\citenamefont {Barausse}\ \emph {et~al.}(2020)\citenamefont
  {Barausse} \emph {et~al.}}]{Barausse2020}%
  \BibitemOpen
  \bibfield  {author} {\bibinfo {author} {\bibfnamefont {E.}~\bibnamefont
  {Barausse}} \emph {et~al.},\ }\bibfield  {title} {\bibinfo {title} {Prospects
  for fundamental physics with lisa},\ }\href
  {https://doi.org/10.1007/s10714-020-02691-1} {\bibfield  {journal} {\bibinfo
  {journal} {General Relativity and Gravitation}\ }\textbf {\bibinfo {volume}
  {52}},\ \bibinfo {pages} {81} (\bibinfo {year} {2020})}\BibitemShut {NoStop}%
\bibitem [{\citenamefont {Piovano}\ \emph {et~al.}(2023)\citenamefont
  {Piovano}, \citenamefont {Maselli},\ and\ \citenamefont
  {Pani}}]{PhysRevD.107.024021}%
  \BibitemOpen
  \bibfield  {author} {\bibinfo {author} {\bibfnamefont {G.~A.}\ \bibnamefont
  {Piovano}}, \bibinfo {author} {\bibfnamefont {A.}~\bibnamefont {Maselli}},\
  and\ \bibinfo {author} {\bibfnamefont {P.}~\bibnamefont {Pani}},\ }\bibfield
  {title} {\bibinfo {title} {Constraining the tidal deformability of
  supermassive objects with extreme mass ratio inspirals and semianalytical
  frequency-domain waveforms},\ }\href
  {https://doi.org/10.1103/PhysRevD.107.024021} {\bibfield  {journal} {\bibinfo
   {journal} {Phys. Rev. D}\ }\textbf {\bibinfo {volume} {107}},\ \bibinfo
  {pages} {024021} (\bibinfo {year} {2023})}\BibitemShut {NoStop}%
\bibitem [{\citenamefont {Hinderer}\ and\ \citenamefont
  {Flanagan}(2008)}]{PhysRevD.78.064028}%
  \BibitemOpen
  \bibfield  {author} {\bibinfo {author} {\bibfnamefont {T.}~\bibnamefont
  {Hinderer}}\ and\ \bibinfo {author} {\bibfnamefont {E.~E.}\ \bibnamefont
  {Flanagan}},\ }\bibfield  {title} {\bibinfo {title} {Two-timescale analysis
  of extreme mass ratio inspirals in kerr spacetime: Orbital motion},\ }\href
  {https://doi.org/10.1103/PhysRevD.78.064028} {\bibfield  {journal} {\bibinfo
  {journal} {Phys. Rev. D}\ }\textbf {\bibinfo {volume} {78}},\ \bibinfo
  {pages} {064028} (\bibinfo {year} {2008})}\BibitemShut {NoStop}%
\bibitem [{\citenamefont {Ramos-Buades}\ \emph {et~al.}(2022)\citenamefont
  {Ramos-Buades}, \citenamefont {van~de Meent}, \citenamefont {Pfeiffer},
  \citenamefont {R\"uter}, \citenamefont {Scheel}, \citenamefont {Boyle},\ and\
  \citenamefont {Kidder}}]{PhysRevD.106.124040}%
  \BibitemOpen
  \bibfield  {author} {\bibinfo {author} {\bibfnamefont {A.}~\bibnamefont
  {Ramos-Buades}}, \bibinfo {author} {\bibfnamefont {M.}~\bibnamefont {van~de
  Meent}}, \bibinfo {author} {\bibfnamefont {H.~P.}\ \bibnamefont {Pfeiffer}},
  \bibinfo {author} {\bibfnamefont {H.~R.}\ \bibnamefont {R\"uter}}, \bibinfo
  {author} {\bibfnamefont {M.~A.}\ \bibnamefont {Scheel}}, \bibinfo {author}
  {\bibfnamefont {M.}~\bibnamefont {Boyle}},\ and\ \bibinfo {author}
  {\bibfnamefont {L.~E.}\ \bibnamefont {Kidder}},\ }\bibfield  {title}
  {\bibinfo {title} {Eccentric binary black holes: Comparing numerical
  relativity and small mass-ratio perturbation theory},\ }\href
  {https://doi.org/10.1103/PhysRevD.106.124040} {\bibfield  {journal} {\bibinfo
   {journal} {Phys. Rev. D}\ }\textbf {\bibinfo {volume} {106}},\ \bibinfo
  {pages} {124040} (\bibinfo {year} {2022})}\BibitemShut {NoStop}%
\bibitem [{\citenamefont {Mino}\ \emph {et~al.}(1997)\citenamefont {Mino},
  \citenamefont {Sasaki}, \citenamefont {Shibata}, \citenamefont {Tagoshi},\
  and\ \citenamefont {Tanaka}}]{10.1143/PTPS.128.1}%
  \BibitemOpen
  \bibfield  {author} {\bibinfo {author} {\bibfnamefont {Y.}~\bibnamefont
  {Mino}}, \bibinfo {author} {\bibfnamefont {M.}~\bibnamefont {Sasaki}},
  \bibinfo {author} {\bibfnamefont {M.}~\bibnamefont {Shibata}}, \bibinfo
  {author} {\bibfnamefont {H.}~\bibnamefont {Tagoshi}},\ and\ \bibinfo {author}
  {\bibfnamefont {T.}~\bibnamefont {Tanaka}},\ }\bibfield  {title} {\bibinfo
  {title} {{Chapter 1. Black Hole Perturbation}},\ }\href
  {https://doi.org/10.1143/PTPS.128.1} {\bibfield  {journal} {\bibinfo
  {journal} {Progress of Theoretical Physics Supplement}\ }\textbf {\bibinfo
  {volume} {128}},\ \bibinfo {pages} {1} (\bibinfo {year} {1997})},\ \Eprint
  {https://arxiv.org/abs/https://academic.oup.com/ptps/article-pdf/doi/10.1143/PTPS.128.1/5438984/128-1.pdf}
  {https://academic.oup.com/ptps/article-pdf/doi/10.1143/PTPS.128.1/5438984/128-1.pdf}
  \BibitemShut {NoStop}%
\bibitem [{\citenamefont {Chatterjee}\ \emph {et~al.}(2023)\citenamefont
  {Chatterjee}, \citenamefont {Mondal},\ and\ \citenamefont
  {Basu}}]{10.1093/mnras/stad3132}%
  \BibitemOpen
  \bibfield  {author} {\bibinfo {author} {\bibfnamefont {S.}~\bibnamefont
  {Chatterjee}}, \bibinfo {author} {\bibfnamefont {S.}~\bibnamefont {Mondal}},\
  and\ \bibinfo {author} {\bibfnamefont {P.}~\bibnamefont {Basu}},\ }\bibfield
  {title} {\bibinfo {title} {{Detectability of gas-riched E/IMRI’s in LISA
  band: Observable signature of transonic accretion flow.}},\ }\href
  {https://doi.org/10.1093/mnras/stad3132} {\bibfield  {journal} {\bibinfo
  {journal} {Monthly Notices of the Royal Astronomical Society}\ ,\ \bibinfo
  {pages} {stad3132}} (\bibinfo {year} {2023})},\ \Eprint
  {https://arxiv.org/abs/https://academic.oup.com/mnras/advance-article-pdf/doi/10.1093/mnras/stad3132/52057839/stad3132.pdf}
  {https://academic.oup.com/mnras/advance-article-pdf/doi/10.1093/mnras/stad3132/52057839/stad3132.pdf}
  \BibitemShut {NoStop}%
\bibitem [{\citenamefont {{Franchini}}\ \emph {et~al.}(2023)\citenamefont
  {{Franchini}}, \citenamefont {{Bonetti}}, \citenamefont {{Lupi}},
  \citenamefont {{Miniutti}}, \citenamefont {{Bortolas}}, \citenamefont
  {{Giustini}}, \citenamefont {{Dotti}}, \citenamefont {{Sesana}},
  \citenamefont {{Arcodia}},\ and\ \citenamefont
  {{Ryu}}}]{2023A&A...675A.100F}%
  \BibitemOpen
  \bibfield  {author} {\bibinfo {author} {\bibfnamefont {A.}~\bibnamefont
  {{Franchini}}}, \bibinfo {author} {\bibfnamefont {M.}~\bibnamefont
  {{Bonetti}}}, \bibinfo {author} {\bibfnamefont {A.}~\bibnamefont {{Lupi}}},
  \bibinfo {author} {\bibfnamefont {G.}~\bibnamefont {{Miniutti}}}, \bibinfo
  {author} {\bibfnamefont {E.}~\bibnamefont {{Bortolas}}}, \bibinfo {author}
  {\bibfnamefont {M.}~\bibnamefont {{Giustini}}}, \bibinfo {author}
  {\bibfnamefont {M.}~\bibnamefont {{Dotti}}}, \bibinfo {author} {\bibfnamefont
  {A.}~\bibnamefont {{Sesana}}}, \bibinfo {author} {\bibfnamefont
  {R.}~\bibnamefont {{Arcodia}}},\ and\ \bibinfo {author} {\bibfnamefont
  {T.}~\bibnamefont {{Ryu}}},\ }\bibfield  {title} {\bibinfo {title}
  {{Quasi-periodic eruptions from impacts between the secondary and a rigidly
  precessing accretion disc in an extreme mass-ratio inspiral system}},\ }\href
  {https://doi.org/10.1051/0004-6361/202346565} {\bibfield  {journal} {\bibinfo
   {journal} {aap}\ }\textbf {\bibinfo {volume} {675}},\ \bibinfo {eid} {A100}
  (\bibinfo {year} {2023})},\ \Eprint {https://arxiv.org/abs/2304.00775}
  {arXiv:2304.00775 [astro-ph.HE]} \BibitemShut {NoStop}%
\bibitem [{\citenamefont {Cole}\ \emph {et~al.}(2023)\citenamefont {Cole},
  \citenamefont {Bertone}, \citenamefont {Coogan}, \citenamefont {Gaggero},
  \citenamefont {Karydas}, \citenamefont {Kavanagh}, \citenamefont {Spieksma},\
  and\ \citenamefont {Tomaselli}}]{Cole2023}%
  \BibitemOpen
  \bibfield  {author} {\bibinfo {author} {\bibfnamefont {P.~S.}\ \bibnamefont
  {Cole}}, \bibinfo {author} {\bibfnamefont {G.}~\bibnamefont {Bertone}},
  \bibinfo {author} {\bibfnamefont {A.}~\bibnamefont {Coogan}}, \bibinfo
  {author} {\bibfnamefont {D.}~\bibnamefont {Gaggero}}, \bibinfo {author}
  {\bibfnamefont {T.}~\bibnamefont {Karydas}}, \bibinfo {author} {\bibfnamefont
  {B.~J.}\ \bibnamefont {Kavanagh}}, \bibinfo {author} {\bibfnamefont
  {T.~F.~M.}\ \bibnamefont {Spieksma}},\ and\ \bibinfo {author} {\bibfnamefont
  {G.~M.}\ \bibnamefont {Tomaselli}},\ }\bibfield  {title} {\bibinfo {title}
  {Distinguishing environmental effects on binary black hole gravitational
  waveforms},\ }\href {https://doi.org/10.1038/s41550-023-01990-2} {\bibfield
  {journal} {\bibinfo  {journal} {Nature Astronomy}\ }\textbf {\bibinfo
  {volume} {7}},\ \bibinfo {pages} {943} (\bibinfo {year} {2023})}\BibitemShut
  {NoStop}%
\bibitem [{\citenamefont {Maselli}\ \emph {et~al.}(2020)\citenamefont
  {Maselli}, \citenamefont {Franchini}, \citenamefont {Gualtieri},\ and\
  \citenamefont {Sotiriou}}]{PhysRevLett.125.141101}%
  \BibitemOpen
  \bibfield  {author} {\bibinfo {author} {\bibfnamefont {A.}~\bibnamefont
  {Maselli}}, \bibinfo {author} {\bibfnamefont {N.}~\bibnamefont {Franchini}},
  \bibinfo {author} {\bibfnamefont {L.}~\bibnamefont {Gualtieri}},\ and\
  \bibinfo {author} {\bibfnamefont {T.~P.}\ \bibnamefont {Sotiriou}},\
  }\bibfield  {title} {\bibinfo {title} {Detecting scalar fields with extreme
  mass ratio inspirals},\ }\href
  {https://doi.org/10.1103/PhysRevLett.125.141101} {\bibfield  {journal}
  {\bibinfo  {journal} {Phys. Rev. Lett.}\ }\textbf {\bibinfo {volume} {125}},\
  \bibinfo {pages} {141101} (\bibinfo {year} {2020})}\BibitemShut {NoStop}%
\bibitem [{\citenamefont {Barsanti}\ \emph {et~al.}(2023)\citenamefont
  {Barsanti}, \citenamefont {Maselli}, \citenamefont {Sotiriou},\ and\
  \citenamefont {Gualtieri}}]{PhysRevLett.131.051401}%
  \BibitemOpen
  \bibfield  {author} {\bibinfo {author} {\bibfnamefont {S.}~\bibnamefont
  {Barsanti}}, \bibinfo {author} {\bibfnamefont {A.}~\bibnamefont {Maselli}},
  \bibinfo {author} {\bibfnamefont {T.~P.}\ \bibnamefont {Sotiriou}},\ and\
  \bibinfo {author} {\bibfnamefont {L.}~\bibnamefont {Gualtieri}},\ }\bibfield
  {title} {\bibinfo {title} {Detecting massive scalar fields with extreme
  mass-ratio inspirals},\ }\href
  {https://doi.org/10.1103/PhysRevLett.131.051401} {\bibfield  {journal}
  {\bibinfo  {journal} {Phys. Rev. Lett.}\ }\textbf {\bibinfo {volume} {131}},\
  \bibinfo {pages} {051401} (\bibinfo {year} {2023})}\BibitemShut {NoStop}%
\bibitem [{\citenamefont {Zhang}\ \emph {et~al.}(2023)\citenamefont {Zhang},
  \citenamefont {Guo}, \citenamefont {Gong},\ and\ \citenamefont
  {Wang}}]{Zhang_2023}%
  \BibitemOpen
  \bibfield  {author} {\bibinfo {author} {\bibfnamefont {C.}~\bibnamefont
  {Zhang}}, \bibinfo {author} {\bibfnamefont {H.}~\bibnamefont {Guo}}, \bibinfo
  {author} {\bibfnamefont {Y.}~\bibnamefont {Gong}},\ and\ \bibinfo {author}
  {\bibfnamefont {B.}~\bibnamefont {Wang}},\ }\bibfield  {title} {\bibinfo
  {title} {Detecting vector charge with extreme mass ratio inspirals onto kerr
  black holes},\ }\href {https://doi.org/10.1088/1475-7516/2023/06/020}
  {\bibfield  {journal} {\bibinfo  {journal} {Journal of Cosmology and
  Astroparticle Physics}\ }\textbf {\bibinfo {volume} {2023}}\bibinfo  {number}
  { (06)},\ \bibinfo {pages} {020}}\BibitemShut {NoStop}%
\bibitem [{\citenamefont {Liang}\ \emph {et~al.}(2023)\citenamefont {Liang},
  \citenamefont {Xu}, \citenamefont {Mai},\ and\ \citenamefont
  {Shao}}]{PhysRevD.107.044053}%
  \BibitemOpen
\bibfield  {number} {  }\bibfield  {author} {\bibinfo {author} {\bibfnamefont
  {D.}~\bibnamefont {Liang}}, \bibinfo {author} {\bibfnamefont
  {R.}~\bibnamefont {Xu}}, \bibinfo {author} {\bibfnamefont {Z.-F.}\
  \bibnamefont {Mai}},\ and\ \bibinfo {author} {\bibfnamefont {L.}~\bibnamefont
  {Shao}},\ }\bibfield  {title} {\bibinfo {title} {Probing vector hair of black
  holes with extreme-mass-ratio inspirals},\ }\href
  {https://doi.org/10.1103/PhysRevD.107.044053} {\bibfield  {journal} {\bibinfo
   {journal} {Phys. Rev. D}\ }\textbf {\bibinfo {volume} {107}},\ \bibinfo
  {pages} {044053} (\bibinfo {year} {2023})}\BibitemShut {NoStop}%
\bibitem [{\citenamefont {Rahman}\ and\ \citenamefont
  {Bhattacharyya}(2023)}]{PhysRevD.107.024006}%
  \BibitemOpen
  \bibfield  {author} {\bibinfo {author} {\bibfnamefont {M.}~\bibnamefont
  {Rahman}}\ and\ \bibinfo {author} {\bibfnamefont {A.}~\bibnamefont
  {Bhattacharyya}},\ }\bibfield  {title} {\bibinfo {title} {Prospects for
  determining the nature of the secondaries of extreme mass-ratio inspirals
  using the spin-induced quadrupole deformation},\ }\href
  {https://doi.org/10.1103/PhysRevD.107.024006} {\bibfield  {journal} {\bibinfo
   {journal} {Phys. Rev. D}\ }\textbf {\bibinfo {volume} {107}},\ \bibinfo
  {pages} {024006} (\bibinfo {year} {2023})}\BibitemShut {NoStop}%
\bibitem [{\citenamefont {Piovano}\ \emph {et~al.}(2020)\citenamefont
  {Piovano}, \citenamefont {Maselli},\ and\ \citenamefont
  {Pani}}]{PhysRevD.102.024041}%
  \BibitemOpen
  \bibfield  {author} {\bibinfo {author} {\bibfnamefont {G.~A.}\ \bibnamefont
  {Piovano}}, \bibinfo {author} {\bibfnamefont {A.}~\bibnamefont {Maselli}},\
  and\ \bibinfo {author} {\bibfnamefont {P.}~\bibnamefont {Pani}},\ }\bibfield
  {title} {\bibinfo {title} {Extreme mass ratio inspirals with spinning
  secondary: A detailed study of equatorial circular motion},\ }\href
  {https://doi.org/10.1103/PhysRevD.102.024041} {\bibfield  {journal} {\bibinfo
   {journal} {Phys. Rev. D}\ }\textbf {\bibinfo {volume} {102}},\ \bibinfo
  {pages} {024041} (\bibinfo {year} {2020})}\BibitemShut {NoStop}%
\bibitem [{\citenamefont {Eda}\ \emph {et~al.}(2013)\citenamefont {Eda},
  \citenamefont {Itoh}, \citenamefont {Kuroyanagi},\ and\ \citenamefont
  {Silk}}]{PhysRevLett.110.221101}%
  \BibitemOpen
  \bibfield  {author} {\bibinfo {author} {\bibfnamefont {K.}~\bibnamefont
  {Eda}}, \bibinfo {author} {\bibfnamefont {Y.}~\bibnamefont {Itoh}}, \bibinfo
  {author} {\bibfnamefont {S.}~\bibnamefont {Kuroyanagi}},\ and\ \bibinfo
  {author} {\bibfnamefont {J.}~\bibnamefont {Silk}},\ }\bibfield  {title}
  {\bibinfo {title} {New probe of dark-matter properties: Gravitational waves
  from an intermediate-mass black hole embedded in a dark-matter minispike},\
  }\href {https://doi.org/10.1103/PhysRevLett.110.221101} {\bibfield  {journal}
  {\bibinfo  {journal} {Phys. Rev. Lett.}\ }\textbf {\bibinfo {volume} {110}},\
  \bibinfo {pages} {221101} (\bibinfo {year} {2013})}\BibitemShut {NoStop}%
\bibitem [{\citenamefont {Li}\ \emph {et~al.}(2022)\citenamefont {Li},
  \citenamefont {Tang},\ and\ \citenamefont {Wu}}]{Li2022}%
  \BibitemOpen
  \bibfield  {author} {\bibinfo {author} {\bibfnamefont {G.-L.}\ \bibnamefont
  {Li}}, \bibinfo {author} {\bibfnamefont {Y.}~\bibnamefont {Tang}},\ and\
  \bibinfo {author} {\bibfnamefont {Y.-L.}\ \bibnamefont {Wu}},\ }\bibfield
  {title} {\bibinfo {title} {Probing dark matter spikes via gravitational waves
  of extreme-mass-ratio inspirals},\ }\href
  {https://doi.org/10.1007/s11433-022-1930-9} {\bibfield  {journal} {\bibinfo
  {journal} {Science China Physics, Mechanics {\&} Astronomy}\ }\textbf
  {\bibinfo {volume} {65}},\ \bibinfo {pages} {100412} (\bibinfo {year}
  {2022})}\BibitemShut {NoStop}%
\bibitem [{\citenamefont {Brito}\ and\ \citenamefont
  {Shah}(2023)}]{PhysRevD.108.084019}%
  \BibitemOpen
  \bibfield  {author} {\bibinfo {author} {\bibfnamefont {R.}~\bibnamefont
  {Brito}}\ and\ \bibinfo {author} {\bibfnamefont {S.}~\bibnamefont {Shah}},\
  }\bibfield  {title} {\bibinfo {title} {Extreme mass-ratio inspirals into
  black holes surrounded by scalar clouds},\ }\href
  {https://doi.org/10.1103/PhysRevD.108.084019} {\bibfield  {journal} {\bibinfo
   {journal} {Phys. Rev. D}\ }\textbf {\bibinfo {volume} {108}},\ \bibinfo
  {pages} {084019} (\bibinfo {year} {2023})}\BibitemShut {NoStop}%
\bibitem [{\citenamefont {Steinhoff}\ and\ \citenamefont
  {Puetzfeld}(2010)}]{PhysRevD.81.044019}%
  \BibitemOpen
  \bibfield  {author} {\bibinfo {author} {\bibfnamefont {J.}~\bibnamefont
  {Steinhoff}}\ and\ \bibinfo {author} {\bibfnamefont {D.}~\bibnamefont
  {Puetzfeld}},\ }\bibfield  {title} {\bibinfo {title} {Multipolar equations of
  motion for extended test bodies in general relativity},\ }\href
  {https://doi.org/10.1103/PhysRevD.81.044019} {\bibfield  {journal} {\bibinfo
  {journal} {Phys. Rev. D}\ }\textbf {\bibinfo {volume} {81}},\ \bibinfo
  {pages} {044019} (\bibinfo {year} {2010})}\BibitemShut {NoStop}%
\bibitem [{\citenamefont {Shen}\ \emph {et~al.}(2023)\citenamefont {Shen},
  \citenamefont {Han} \emph {et~al.}}]{PhysRevD.108.064015}%
  \BibitemOpen
  \bibfield  {author} {\bibinfo {author} {\bibfnamefont {P.}~\bibnamefont
  {Shen}}, \bibinfo {author} {\bibfnamefont {W.-B.}\ \bibnamefont {Han}}, \emph
  {et~al.},\ }\bibfield  {title} {\bibinfo {title} {Influence of mass-ratio
  corrections in extreme-mass-ratio inspirals for testing general relativity},\
  }\href {https://doi.org/10.1103/PhysRevD.108.064015} {\bibfield  {journal}
  {\bibinfo  {journal} {Phys. Rev. D}\ }\textbf {\bibinfo {volume} {108}},\
  \bibinfo {pages} {064015} (\bibinfo {year} {2023})}\BibitemShut {NoStop}%
\bibitem [{\citenamefont {{Teukolsky}}(1973)}]{1973ApJ...185..635T}%
  \BibitemOpen
  \bibfield  {author} {\bibinfo {author} {\bibfnamefont {S.~A.}\ \bibnamefont
  {{Teukolsky}}},\ }\bibfield  {title} {\bibinfo {title} {{Perturbations of a
  Rotating Black Hole. I. Fundamental Equations for Gravitational,
  Electromagnetic, and Neutrino-Field Perturbations}},\ }\href
  {https://doi.org/10.1086/152444} {\bibfield  {journal} {\bibinfo  {journal}
  {apj}\ }\textbf {\bibinfo {volume} {185}},\ \bibinfo {pages} {635} (\bibinfo
  {year} {1973})}\BibitemShut {NoStop}%
\bibitem [{\citenamefont {{Press}}\ and\ \citenamefont
  {{Teukolsky}}(1973)}]{1973ApJ...185..649P}%
  \BibitemOpen
  \bibfield  {author} {\bibinfo {author} {\bibfnamefont {W.~H.}\ \bibnamefont
  {{Press}}}\ and\ \bibinfo {author} {\bibfnamefont {S.~A.}\ \bibnamefont
  {{Teukolsky}}},\ }\bibfield  {title} {\bibinfo {title} {{Perturbations of a
  Rotating Black Hole. II. Dynamical Stability of the Kerr Metric}},\ }\href
  {https://doi.org/10.1086/152445} {\bibfield  {journal} {\bibinfo  {journal}
  {apj}\ }\textbf {\bibinfo {volume} {185}},\ \bibinfo {pages} {649} (\bibinfo
  {year} {1973})}\BibitemShut {NoStop}%
\bibitem [{\citenamefont {{Teukolsky}}\ and\ \citenamefont
  {{Press}}(1974)}]{1974ApJ...193..443T}%
  \BibitemOpen
  \bibfield  {author} {\bibinfo {author} {\bibfnamefont {S.~A.}\ \bibnamefont
  {{Teukolsky}}}\ and\ \bibinfo {author} {\bibfnamefont {W.~H.}\ \bibnamefont
  {{Press}}},\ }\bibfield  {title} {\bibinfo {title} {{Perturbations of a
  rotating black hole. III. Interaction of the hole with gravitational and
  electromagnetic radiation.}},\ }\href {https://doi.org/10.1086/153180}
  {\bibfield  {journal} {\bibinfo  {journal} {apj}\ }\textbf {\bibinfo {volume}
  {193}},\ \bibinfo {pages} {443} (\bibinfo {year} {1974})}\BibitemShut
  {NoStop}%
\bibitem [{\citenamefont {Tulczyjew}(1959)}]{tulczyjew1959motion}%
  \BibitemOpen
  \bibfield  {author} {\bibinfo {author} {\bibfnamefont {W.}~\bibnamefont
  {Tulczyjew}},\ }\bibfield  {title} {\bibinfo {title} {Motion of multipole
  particles in general relativity theory},\ }\href@noop {} {\bibfield
  {journal} {\bibinfo  {journal} {Acta Phys. Pol}\ }\textbf {\bibinfo {volume}
  {18}},\ \bibinfo {pages} {94} (\bibinfo {year} {1959})}\BibitemShut {NoStop}%
\bibitem [{\citenamefont {Isaacson}(1968)}]{PhysRev.166.1272}%
  \BibitemOpen
  \bibfield  {author} {\bibinfo {author} {\bibfnamefont {R.~A.}\ \bibnamefont
  {Isaacson}},\ }\bibfield  {title} {\bibinfo {title} {Gravitational radiation
  in the limit of high frequency. ii. nonlinear terms and the effective stress
  tensor},\ }\href {https://doi.org/10.1103/PhysRev.166.1272} {\bibfield
  {journal} {\bibinfo  {journal} {Phys. Rev.}\ }\textbf {\bibinfo {volume}
  {166}},\ \bibinfo {pages} {1272} (\bibinfo {year} {1968})}\BibitemShut
  {NoStop}%
\bibitem [{\citenamefont {Hawking}\ and\ \citenamefont
  {Hartle}(1972)}]{Hawking1972}%
  \BibitemOpen
  \bibfield  {author} {\bibinfo {author} {\bibfnamefont {S.~W.}\ \bibnamefont
  {Hawking}}\ and\ \bibinfo {author} {\bibfnamefont {J.~B.}\ \bibnamefont
  {Hartle}},\ }\bibfield  {title} {\bibinfo {title} {Energy and angular
  momentum flow into a black hole},\ }\href
  {https://doi.org/10.1007/BF01645515} {\bibfield  {journal} {\bibinfo
  {journal} {Communications in Mathematical Physics}\ }\textbf {\bibinfo
  {volume} {27}},\ \bibinfo {pages} {283} (\bibinfo {year} {1972})}\BibitemShut
  {NoStop}%
\bibitem [{\citenamefont {Bonga}\ \emph {et~al.}(2019)\citenamefont {Bonga},
  \citenamefont {Yang},\ and\ \citenamefont {Hughes}}]{PhysRevLett.123.101103}%
  \BibitemOpen
  \bibfield  {author} {\bibinfo {author} {\bibfnamefont {B.}~\bibnamefont
  {Bonga}}, \bibinfo {author} {\bibfnamefont {H.}~\bibnamefont {Yang}},\ and\
  \bibinfo {author} {\bibfnamefont {S.~A.}\ \bibnamefont {Hughes}},\ }\bibfield
   {title} {\bibinfo {title} {Tidal resonance in extreme mass-ratio
  inspirals},\ }\href {https://doi.org/10.1103/PhysRevLett.123.101103}
  {\bibfield  {journal} {\bibinfo  {journal} {Phys. Rev. Lett.}\ }\textbf
  {\bibinfo {volume} {123}},\ \bibinfo {pages} {101103} (\bibinfo {year}
  {2019})}\BibitemShut {NoStop}%
\bibitem [{\citenamefont {Vallisneri}(2008)}]{PhysRevD.77.042001}%
  \BibitemOpen
  \bibfield  {author} {\bibinfo {author} {\bibfnamefont {M.}~\bibnamefont
  {Vallisneri}},\ }\bibfield  {title} {\bibinfo {title} {Use and abuse of the
  fisher information matrix in the assessment of gravitational-wave
  parameter-estimation prospects},\ }\href
  {https://doi.org/10.1103/PhysRevD.77.042001} {\bibfield  {journal} {\bibinfo
  {journal} {Phys. Rev. D}\ }\textbf {\bibinfo {volume} {77}},\ \bibinfo
  {pages} {042001} (\bibinfo {year} {2008})}\BibitemShut {NoStop}%
\bibitem [{\citenamefont {Hughes}(2000)}]{PhysRevD.61.084004}%
  \BibitemOpen
  \bibfield  {author} {\bibinfo {author} {\bibfnamefont {S.~A.}\ \bibnamefont
  {Hughes}},\ }\bibfield  {title} {\bibinfo {title} {Evolution of circular,
  nonequatorial orbits of kerr black holes due to gravitational-wave
  emission},\ }\href {https://doi.org/10.1103/PhysRevD.61.084004} {\bibfield
  {journal} {\bibinfo  {journal} {Phys. Rev. D}\ }\textbf {\bibinfo {volume}
  {61}},\ \bibinfo {pages} {084004} (\bibinfo {year} {2000})}\BibitemShut
  {NoStop}%
\end{thebibliography}
\providecommand{\noopsort}[1]{}\providecommand{\singleletter}[1]{#1}%

\appendix
\section{Proof}\label{appendix: proof}
In Boyer-Lindquist coordinates, the stress-energy tensor present in Eq.~(\ref{eq: stress-energy tensor}) can be transformed to the following form:
\begin{equation}
    T^{\alpha\beta}=\int \left\{t^{\alpha\beta}_\text{lc}\delta(x-z(\lambda))+\partial_\gamma[t^{\gamma\alpha\beta}_\text{lc}\delta(x-z(\lambda))]+\partial_\delta\partial_\gamma[t^{\delta\gamma\alpha\beta}_\text{lc}\delta(x-z(\lambda))]+ ... \right\}d\lambda
\end{equation}
where $(t^{\alpha\beta}_\text{lc},t^{\gamma\alpha\beta}_\text{lc},t^{\delta\gamma\alpha\beta}_\text{lc},...)$ are all the linear combination of $(t^{\alpha\beta},t^{\gamma\alpha\beta},t^{\delta\gamma\alpha\beta},...)$, we denote 
\begin{equation}
    (t^{\alpha\beta}_\text{lc},t^{\gamma\alpha\beta}_\text{lc},...)=L_0(t^{\alpha\beta},t^{\gamma\alpha\beta},...),
\end{equation}
and it's worth pointing out that the coefficients of this linear combination are independent of $t$ and $\phi$.

The background spacetime is also described by the Kerr metric in Boyer-Lindquist coordinates,
\begin{equation}
    ds^2=-dt^2+\Sigma(\Delta^{-1}dr^2+d\theta^2)+(r^2+a^2)\sin^2{\theta}d\phi^2+2Mr/\Sigma(a\sin^2{\theta}-dt)^2,
\end{equation}
where $\Delta=r^2-2Mr+a^2$, $\Sigma=r^2+a^2\cos^2{\theta}$.
Teukolsky formalism decomposed $\psi_4$ using the Newman-Penrose tetrad basis, which are
\begin{equation}
    \begin{aligned}
        l^\alpha=&\left(r^2+a^2,\Delta,0,a\right)/\Delta,\\
        n^\alpha=&\left((r^2+a^2),-\Delta,0,a\right)/(2\Sigma),\\
        m^\alpha=&\left(ia\sin\theta,0,1,\frac{i}{\sin\theta}\right)/(\sqrt{2}(r+ia\cos\theta)).
    \end{aligned}
\end{equation}
The radial Teukolsky equation is given by
\begin{equation}
    \Delta^2\frac{d}{dr}\left(\frac{1}{\Delta}\frac{dR_{lm\omega}}{dr}\right)-V(r)R_{lm\omega}(r)=\mathcal{T}_{lm\omega},
\end{equation}
where 
\begin{align}
    V(r)=&-\frac{K^2+4i(r-1)K}{\Delta}+8i\omega r+\lambda_{lm\omega},\\
    K=&(r^2+a^2)\omega-a\omega,
\end{align}
$\lambda_{lm\omega}$ is the eigenvalues of angular Teukolsky equation. For the source term $\mathcal{T}_{lm\omega}$ we adopt the form given by Ref.~\cite{PhysRevD.102.024041},
\begin{equation}
    \mathcal{T}_{lm\omega}=\int dtd\theta d\phi e^{i(\omega t-m\phi)}(\mathcal{T}_{nn}+\mathcal{T}_{n\bar{m}}+\mathcal{T}_{\bar{m}\bar{m}}),
\end{equation}
where
\begin{align}
    \mathcal{T}_{nn}=&f^{0nn}_{\mu\nu}T^{\mu\nu},\\
    \mathcal{T}_{n\bar{m}}=&f^{0n\bar{m}}_{\mu\nu}T^{\mu\nu}+\partial_r(f^{1n\bar{m}}_{\mu\nu}T^{\mu\nu}),\\
    \mathcal{T}_{\bar{m}\bar{m}}=&f^{0\bar{m}\bar{m}}_{\mu\nu}T^{\mu\nu}+\partial_r(f^{1\bar{m}\bar{m}}_{\mu\nu}T^{\mu\nu})+\partial_r^2(f^{2\bar{m}\bar{m}}_{\mu\nu}T^{\mu\nu}),
\end{align}
with
\begin{align}
    f^{0nn}_{\mu\nu}=&-\frac{2\sin\theta}{\Delta^2\rho^3\bar{\rho}}(\mathcal{L}^{\dagger}_1-2ia\sin\theta)[\mathcal{L}^{\dagger}_2[S^{a\omega}_{lm}]]n_{\mu}n_{\nu},\label{eq: f0nn}\\
    f^{0n\bar{m}}_{\mu\nu}=&\frac{4\sin\theta}{\sqrt{2}\rho^3\Delta}\left(\left(\frac{iK}{\Delta}+\rho+\bar{\rho}\right)\mathcal{L}^{\dagger}_2-a\sin\theta\frac{K}{\Delta}(\bar{\rho}-\rho)\right)[S^{a\omega}_{lm}]\left(\frac{n_{\mu}\bar{m}_{\nu}+n_{\nu}\bar{m}_{\mu}}{2}\right),\\
    f^{1n\bar{m}}_{\mu\nu}=&\frac{4\sin\theta}{\sqrt{2}\rho^3\Delta}(\mathcal{L}^{\dagger}_2+ia\sin\theta(\bar{\rho}-\rho))[S^{a\omega}_{lm}]\left(\frac{n_{\mu}\bar{m}_{\nu}+n_{\nu}\bar{m}_{\mu}}{2}\right),\\
    f^{0\bar{m}\bar{m}}_{\mu\nu}=&\frac{\bar{\rho}}{\rho^3}\left(\frac{d}{dr}\left(\frac{iK}{\Delta}\right)-2\rho\frac{iK}{\Delta}+\frac{K^2}{\Delta^2}\right)S^{a\omega}_{lm}\bar{m}_{\mu}\bar{m}_{\nu},\\
    f^{1\bar{m}\bar{m}}_{\mu\nu}=&-2\left(\frac{\bar{\rho}}{\rho^2}+\frac{i\bar{\rho}K}{\rho^3\Delta}\right)S^{a\omega}_{lm}\bar{m}_{\mu}\bar{m}_{\nu},\\
    f^{2\bar{m}\bar{m}}_{\mu\nu}=&-\frac{\bar{\rho}}{\rho^3}S^{a\omega}_{lm}\bar{m}_{\mu}\bar{m}_{\nu}.\label{eq: f2mm}
\end{align}
The function given in Eq.~(\ref{eq: f0nn})-(\ref{eq: f2mm}) only depend $\omega$, $m$, $r$, and $\theta$. Following Ref.~\cite{PhysRevD.61.084004}, the asymptotic amplitude $Z^{H,\infty}_{lm\omega}$ are
\begin{equation}
    \begin{aligned}
        Z^{\infty,H}_{lm\omega}=&C^{\infty,H}_{lm\omega}\int^{\infty}_{r^+}dr\frac{R^{up,in}_{lm\omega}\mathcal{T}_{lm\omega}}{\Delta^2}\\
        =&C^{\infty,H}_{lm\omega}\int^{\infty}_{r^+}dr\int dtd\theta d\phi e^{i(\omega t-m\phi)}(\mathcal{T}_{nn}+\mathcal{T}_{n\bar{m}}+\mathcal{T}_{\bar{m}\bar{m}})R^{up,in}_{lm\omega}\\
        =&C^{\infty,H}_{lm\omega}\int^{\infty}_{r^+}dr\int dtd\theta d\phi e^{i(\omega t-m\phi)}(f^{0}_{\mu\nu}T^{\mu\nu}+\partial_r(f^{1}_{\mu\nu}T^{\mu\nu})+\partial_r^2(f^{2}_{\mu\nu}T^{\mu\nu})R^{up,in}_{lm\omega}
    \end{aligned}
\end{equation}
where
\begin{align}
    f^{0}_{\mu\nu}=&f^{0nn}_{\mu\nu}+f^{0n\bar{m}}_{\mu\nu}+f^{0\bar{m}\bar{m}}_{\mu\nu},\\
    f^{1}_{\mu\nu}=&f^{1n\bar{m}}_{\mu\nu}+f^{1\bar{m}\bar{m}}_{\mu\nu},\\
    f^{2}_{\mu\nu}=&f^{2\bar{m}\bar{m}}_{\mu\nu}.
\end{align}
We will introduce an equation about the Dirac $\delta$ function for the following calculation. If $f$ and $g$ are smooth functions, we have
\begin{equation}
    \int f(x) \partial_x^n (g(x)\delta(x-x_0)) dx=(-1)^ng(x_0)\partial_x^nf(x)|_{x=x_0}.
\end{equation}
Thus, we have
\begin{equation}
    \begin{aligned}
        Z^{\infty,H}_{lm\omega}=&C^{\infty,H}_{lm\omega}\int^{\infty}_{r^+}dr\int dtd\theta d\phi e^{i(\omega t-m\phi)}(f^{0}_{\mu\nu}T^{\mu\nu}+\partial_r(f^{1}_{\mu\nu}T^{\mu\nu})+\partial_r^2(f^{2}_{\mu\nu}T^{\mu\nu}))R^{up,in}_{lm\omega}\\
        =&C^{\infty,H}_{lm\omega}\int^{\infty}_{r^+}dr\int dtd\theta d\phi e^{i(\omega t-m\phi)}(f^{0}_{\mu\nu}R^{up,in}_{lm\omega}-f^{1}_{\mu\nu}\partial_rR^{up,in}_{lm\omega}+f^{2}_{\mu\nu}\partial_r^2R^{up,in}_{lm\omega})T^{\mu\nu}.\\
    \end{aligned}
\end{equation}
Let $h_{\mu\nu}(r,\theta)=(f^{0}_{\mu\nu}R^{up,in}_{lm\omega}-f^{1}_{\mu\nu}\partial_rR^{up,in}_{lm\omega}+f^{2}_{\mu\nu}\partial_r^2R^{up,in}_{lm\omega})$,
\begin{equation}
    \begin{aligned}
        Z^{\infty,H}_{lm\omega}=&C^{\infty,H}_{lm\omega}\int^{\infty}_{r^+}dr\int dtd\theta d\phi e^{i(\omega t-m\phi)}h_{\alpha\beta}(r,\theta)\int \left\{t^{\alpha\beta}_\text{lc}\delta(x-z(\lambda))+\partial_\gamma[t^{\gamma\alpha\beta}_\text{lc}\delta(x-z(\lambda))]+ ... \right\}d\lambda\\
        =&C^{\infty,H}_{lm\omega}\int d\lambda(t^{\alpha\beta}_\text{lc}(e^{i(\omega t-m\phi)}h_{\alpha\beta})+t^{\gamma\alpha\beta}_\text{lc}\partial_\gamma (e^{i(\omega t-m\phi)}h_{\alpha\beta})+...).
    \end{aligned}
\end{equation}
Because of $\partial_t^n e^{i\omega t}=(i\omega)^ne^{i\omega t}$ and $\partial_\phi^n e^{-im\phi}=(-im)^ne^{-im\phi}$, we have
\begin{equation}
    \begin{aligned}
        Z^{\infty,H}_{lm\omega}=&C^{\infty,H}_{lm\omega}\int d\lambda e^{i(\omega t-m\phi)}L_1(t^{\alpha\beta}_\text{lc},t^{\gamma\alpha\beta}_\text{lc},t^{\delta\gamma\alpha\beta}_\text{lc},...)\\
        =&C^{\infty,H}_{lm\omega}\int dt\left(\frac{dt}{d\lambda}\right)^{-1}e^{i(\omega t-m\phi)}L_1L_0(t^{\alpha\beta},t^{\gamma\alpha\beta},t^{\delta\gamma\alpha\beta},...),
    \end{aligned}
\end{equation}
where $L_1$ is a linear operator which only dependent of $\omega$, $m$, $r$, and $\theta$.
If $\frac{dt}{d\lambda}$ and $(t^{\alpha\beta},t^{\gamma\alpha\beta},t^{\delta\gamma\alpha\beta},...)$ independent of $t$ or their time scale of the variation is much larger than the orbital time scale, and $r=r_0$, $\phi=\Omega_0 t$ for circular motion, we result that
\begin{equation}
    \begin{aligned}
        Z^{\infty,H}_{lm\omega}=2\pi \delta(\omega-m\Omega_0)C^{\infty,H}_{lm\omega} L_1L_0(t^{\alpha\beta},t^{\gamma\alpha\beta},t^{\delta\gamma\alpha\beta},...)/v^t.
    \end{aligned}
\end{equation}
This is exactly the form of Eq.~(\ref{eq: decoupling amplitude}), and the proof is completed.
\end{document}